\documentclass[]{aa}
\usepackage{graphicx}
\usepackage{subcaption}
\usepackage{mwe}
\usepackage{txfonts}
\usepackage{hyperref}
\usepackage{url}
\usepackage{xcolor}
\hypersetup{
colorlinks,
linkcolor={red!80!black},
citecolor={blue!80!black},
urlcolor={blue!80!black}
}

\def\cm3{cm$^{-3}$}
\def\kms{km~s$^{-1}$}

\def\rsun{R$_{\odot}$}

\def\msun{M$_{\odot}$}

\def\one{\ts {\,\sc i}}
\def\two{\ts {\,\sc ii}}
\def\three{\ts {\,\sc iii}}
\def\four{\ts {\,\sc iv}}
\def\five{\ts {\sc v}}
\def\six{\ts {\sc vi}}
\def\beq{\begin{equation}}
\def\eeq{\end{equation}}

\def\lesssim{\mathrel{\hbox{\rlap{\hbox{\lower4pt\hbox{$\sim$}}}\hbox{$<$}}}}
\def\gtrsim{\mathrel{\hbox{\rlap{\hbox{\lower4pt\hbox{$\sim$}}}\hbox{$>$}}}}

\def\one{{\,\sc i}}
\def\two{{\,\sc ii}}
\def\three{{\,\sc iii}}
\def\four{{\,\sc iv}}
\def\five{{\sc v}}
\def\six{{\sc vi}}

\newcommand{\code}[1]{\texttt{#1}}

\newcommand{\voned}{\code{V1D}}

\newcommand{\cmfgen}{\code{CMFGEN}}

\newcommand{\arepo}{\code{AREPO}}

\def\ergs{erg\,s$^{-1}$}

\def\aj{AJ}
\def\pasp{PASP}

\def\apj{ApJ}
\def\apjs{ApJS}
\def\apjl{ApJL}
\def\aap{A\&A}
\def\araa{ARA\&A}

\def\mnras{MNRAS}
\def\nat{Nature}

\def\nad{Na\one\,$\lambda\lambda\,5896,5890$}
\def\mgiidoub{Mg\two\,$\lambda\lambda$\,$2795,\,2802$}

\begin{document}

   \title{The look of high-velocity red-giant star collisions}
   \titlerunning{Red-giant star collisions}

\author{
Luc Dessart,\inst{\ref{inst1}}
Taeho Ryu,\inst{\ref{inst2},\ref{inst3}}
Pau Amaro Seoane,\inst{\ref{inst4},\ref{inst5},\ref{inst6},\ref{inst7}}
and Andrew~M.~Taylor\inst{\ref{inst8}}
}

\institute{
Institut d'Astrophysique de Paris, CNRS-Sorbonne Universit\'e, 98 bis boulevard Arago, F-75014 Paris, France\label{inst1}
\and
The Max Planck Institute for Astrophysics, Karl-Schwarzschild-Str. 1, Garching, D-85748, Germany\label{inst2}
\and
Physics and Astronomy Department, Johns Hopkins University, Baltimore, MD 21218, USA\label{inst3}
\and
Universitad Polit\`ecnica de Val\`encia, Spain\label{inst4}
\and
Max Planck Institute for Extraterrestrial Physics, Garching, Germany\label{inst5}
\and
Higgs Centre for Theoretical Physics, Edinburgh, UK\label{inst6}
\and
Kavli Institute for Astronomy and Astrophysics, Beijing 100871, China\label{inst7}
\and
DESY, Zeuthen, Germany\label{inst8}
  }

   \date{Received; accepted}

  \abstract{
High-velocity stellar collisions driven by a supermassive black hole (BH) or BH-driven disruptive collisions, in dense, nuclear clusters can rival the energetics of supergiant star explosions following gravitational collapse of their iron core. Here, starting from a sample of red-giant star collisions simulated with the hydrodynamics code \arepo, we generate photometric and spectroscopic observables using the nonlocal thermodynamic equilibrium time-dependent radiative transfer code \cmfgen. Collisions from more extended giants or stronger collisions (higher velocity or smaller impact parameter) yield bolometric luminosities on the order of 10$^{43}$\,\ergs\ at 1\,d, evolving on a timescale of a week to a bright plateau at $\sim$\,10$^{41}$\,\ergs, before plunging precipitously after 20--40\,d at the end of the optically-thick phase. This luminosity falls primarily in the UV in the first days, thus when it is at its maximum, and shifts to the optical thereafter.  Collisions at lower velocity or from less extended stars produce ejecta that are fainter but may remain optically thick for up to 40\,d if they have a small expansion rate. These collision debris show a similar spectral evolution as that observed or modeled for blue-supergiant star explosions of massive stars, differing only in the more rapid transition to the nebular phase. Such BH-driven disruptive collisions should be detectable by high-cadence surveys in the UV like ULTRASAT. 
}

   \maketitle

\section{Introduction}

In astrophysical environments that have a sufficiently large stellar density, such as globular clusters or the central regions of galaxies (aka nuclear clusters), stars may collide with each other \citep{HillsDay1976,DaleDavies2006}. Such collisions can be significantly more destructive in nuclear clusters than in globular clusters, primarily because of the greater velocity of stars in the former. In globular clusters, the collision velocity is on the order of the velocity dispersion, which is about 10--15\,\kms\ \citep{Cohen+1983}. In nuclear clusters, this collision velocity is in contrast much greater. It is determined by the local Keplerian speed around the central supermassive black hole (BH), thus on the order of 1000\,\kms\,$(M_{\rm BH}/10^{7}M_{\odot})^{1/2}(r/0.05{\rm pc})^{-1/2}$, where $M_{\rm BH}$ is the mass of the BH and $r$ is the distance from the BH. In such cases, the kinetic energy of the collision exceeds the binding energy of each star, resulting in their complete destruction, the formation of a homologously expanding debris, and a burst of radiation analogous to what is routinely observed in supernovae (SNe; \citealt{ryu_collisions_23}). The rate of such BH-driven destructive collisions (BDCs) has been estimated to range from $10^{-4}$ to $10^{-9}$ yr$^{-1}$ galaxy$^{-1}$ for main-sequence stars \citep{Rose+2020,AmaroSeoane2023, Rose+2023, BalbergYassur2023}. This rate depends on various factors, including the distance from the supermassive BH, the influx of stars into the nuclear center, and the core's depletion rate (the latter refers to the rate at which the stellar density in the core decreases because of tidal-disruption events, stellar collisions, and stellar ejections via multi-body interactions). However, \citet{AmaroSeoane2023} recently showed that the event rate must be dominated by high-velocity events, both for main-sequence and red-giant stars. In the case of red-giant stars, due to their larger cross-section, this estimated event rate is even larger, reaching tens of such collisions per year within a volume of radius 100\,Mpc. In the case of main-sequence stars, this number is estimated to be of a few per year in the same volume.

In spite of the potential for detecting such events, the observables of BDCs have not been extensively studied. Recently, using the moving-mesh hydrodynamics code {\small AREPO} \citep{Arepo,Arepo2,ArepoHydro}, \citet{ryu_collisions_23} conducted simulations of collisions between 1\,\msun\ giants with a surface radius $R_{\star}$ in the range from 10 to 100\,\rsun\ and collision (or relative) velocities $v_{\rm rel}$ of 2500 up to 10000\,\kms. Their simulations revealed that for $v_{\rm rel}\gtrsim$\,2500\,\kms\ and impact parameters $b\leq$\,0.8\,$R_{\star}$, the collisions produce a quasi-spherical, supersonic, and homologously expanding debris. Additionally, by considering the radiation energy stored in that ejecta and the local cooling time, they estimated that the peak luminosity occurs essentially at the moment of collision, reaching values as high as 10$^{42}$--10$^{44}$\,\ergs\ depending on the collision parameters. The luminosity subsequently decreases by an order of magnitude during the first week after the collision and stays relatively constant thereafter. The maximum photospheric temperature can reach 10$^5$\,K, which corresponds to a spectral energy distribution (SED) that peaks in the far-UV, and gradually declines over the course of a month down to several $10^{3}$K, which corresponds to an SED that peaks at optical wavelengths. Such properties are in agreement with the analytical work of \citet{AmaroSeoane2023}, which was further expanded in \citet{amaro_col_23b} to address the additional observables investigated in the numerical simulations of \citet{ryu_collisions_23}. The results from this second, analytical work show a good agreement with the numerical predictions. The high luminosity of BDCs suggests that they should be detectable by optical/UV transient surveys such as ZTF \citep{ZTF},\footnote{https://www.ztf.caltech.edu} ASAS-SN \citep{ASASSN},\footnote{https://www.astronomy.ohio-state.edu/asassn} LSST \citep{LSST},\footnote{https://www.lsst.org} or ULTRASAT \citep{ULTRASAT}.\footnote{https://www.weizmann.ac.il/ultrasat}

\begin{table*}
\caption{
Properties of our model set at the start of the \cmfgen\ simulations at about 1\,d after collision. For
each model, we list the total mass and the surface radius of each star, followed by their relative velocity and
    impact parameter at collision, and finally some properties of the ejecta that we model with \cmfgen\ (i.e.
    the ejecta mass, its kinetic energy, and the total stored radiative energy within the ejecta at 1\,d).
    Numbers in parenthesis represent powers of ten.
\label{tab_init}
}
\begin{center}
\begin{tabular}{
l@{\hspace{2mm}}c@{\hspace{2mm}}c@{\hspace{2mm}}
c@{\hspace{2mm}}c@{\hspace{2mm}}c@{\hspace{2mm}}
c@{\hspace{2mm}}c@{\hspace{2mm}}c@{\hspace{2mm}}
c@{\hspace{2mm}}
}
\hline
               Model   & $M_{\star,1}$&$R_{\star,1}$&$M_{\star,2}$&$R_{\star,2}$&$V_{\rm rel}$&       b   &$M_{\rm ej}$  &$E_{\rm kin}$  & $E_{\rm rad}$ \\
                       &   [\msun]   &   [\rsun] &  [\msun]   &   [\rsun] &   [\kms]    &[$R_\star$]& [\msun]     &    [erg]     &   [erg]      \\
\hline
    rs100/v1e4/b0p04   &     0.9     &   100.0   &     0.9     &    100.0  &    10000   &    0.04   &    1.19    &  2.88(50)    &  1.65(49)     \\
     rs50/v1e4/b0p04   &     1.0     &    50.0   &     1.0     &     50.0  &    10000   &    0.04   &    1.36    &  3.42(50)    &  9.36(48)     \\
     rs20/v1e4/b0p04   &     1.0     &    20.0   &     1.0     &     20.0  &    10000   &    0.04   &    1.46    &  3.72(50)    &  3.91(48)     \\
     rs10/v1e4/b0p04   &     1.0     &    10.0   &     1.0     &     10.0  &    10000   &    0.04   &    1.59    &  4.04(50)    &  1.68(48)     \\
      rs10/v1e4/b0p4   &     1.0     &    10.0   &     1.0     &     10.0  &    10000   &    0.40   &    1.59    &  4.02(50)    &  1.39(48)     \\
      rs10/v1e4/b0p8   &     1.0     &    10.0   &     1.0     &     10.0  &    10000   &    0.80   &    1.59    &  4.01(50)    &  7.65(47)     \\
     rs10/v5e3/b0p04   &     1.0     &    10.0   &     1.0     &     10.0  &     5000   &    0.04   &    1.60    &  1.02(50)    &  7.26(47)     \\
   rs10/v2p5e3/b0p04   &     1.0     &    10.0   &     1.0     &     10.0  &     2500   &    0.04   &    1.60    &  2.58(49)    &  1.20(47)     \\
\hline
\end{tabular}
\end{center}
\end{table*}

\begin{figure}
\includegraphics[width=\hsize]{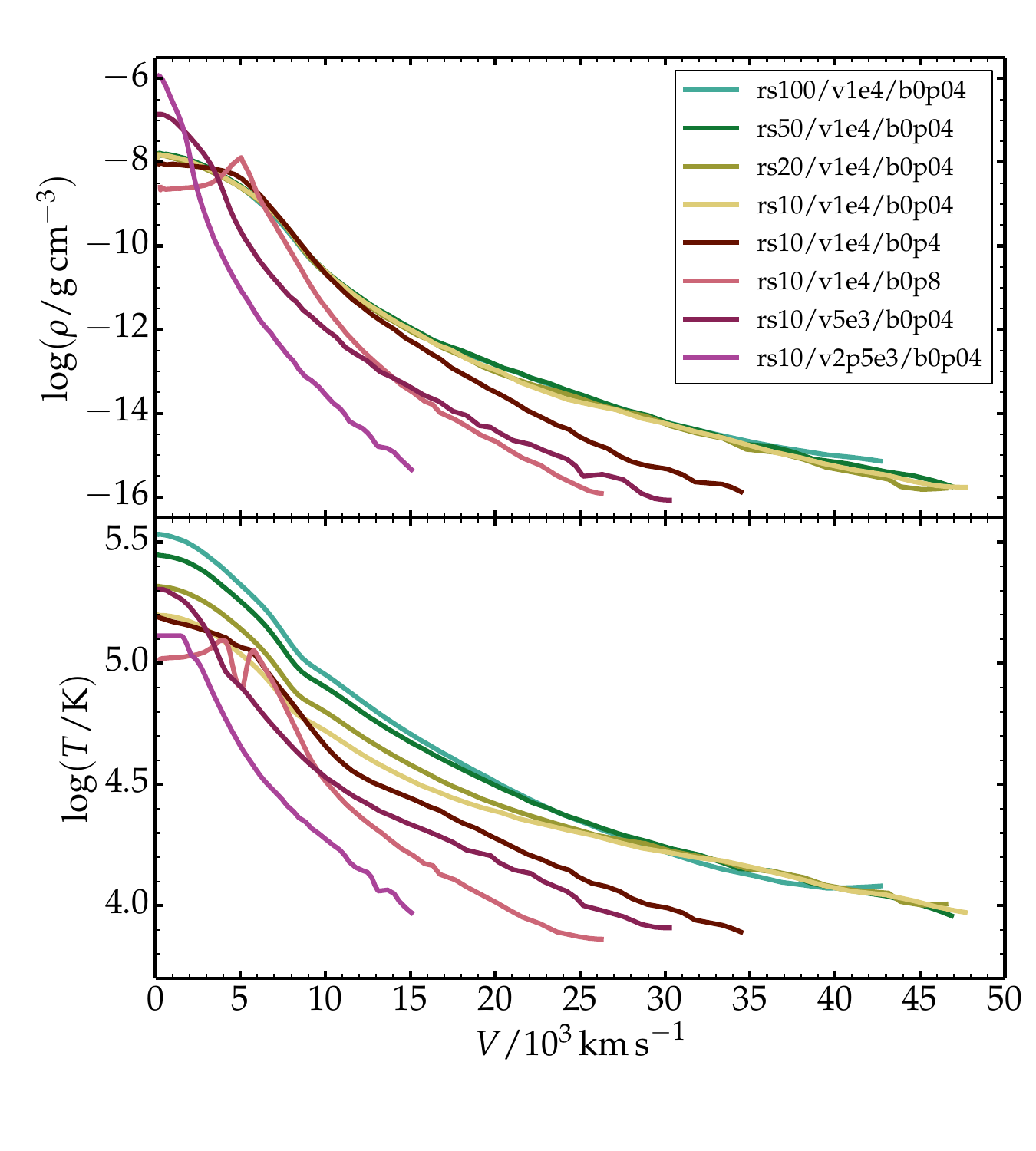}
   \caption{Initial ejecta properties for our model set of high-velocity collisions of red giants. We show the density (top) and the temperature (bottom) for the models at the time of remapping from \arepo\ to \cmfgen, which is at about 1\,d in all cases. Additional properties are given in Table~\ref{tab_init}.}
\label{fig_init}
\end{figure}

The radiative signatures of BDCs inferred from hydrodynamics simulations in \citet{ryu_collisions_23} are only rough estimates. A fixed ionization level was assumed, which ignores the expected recombination of the gas as it cools from 10$^5$ down to several 10$^3$\,K. The adopted fixed ionization level overestimates the duration of the optically-thick, photospheric phase, and consequently underestimates the luminosity at the corresponding epochs. Finally, these estimates remain elusive about the color evolution and the spectral properties of  BDCs. Here, we improve the robustness of these predictions by performing 1-D nonlocal thermodynamic equilibrium (NLTE) time-dependent radiative-transfer calculations for the BDC simulations of  \citet{ryu_collisions_23} using \cmfgen\ \citep{HD12}. The rapid transition to a homologous flow and the quasi-spherical material distribution of the debris obtained in the hydrodynamics simulations supports the approach with \cmfgen, which ignores dynamics and assumes spherical symmetry. The benefit is, however, the state-of-the-art treatment of the radiative transfer, at the same level of sophistication as employed for SNe (see, for example, \citealt{HD19}).

In the next section, we present the numerical approach with \cmfgen, including a description of the mapping procedure from the 3D \arepo\ simulations to the 1-D ejecta we start with at 1\,d after collision of the two stars. Section~\ref{sect_res} describes the results from the \cmfgen\ simulations, with a description of the photometric, spectroscopic, as well as the gas properties for our model set. In section~\ref{sect_conc}, we present our conclusions.

 \section{Numerical approach}

The simulations presented in this work are analogous to those performed for Type II SN ejecta with the code \cmfgen\ \citep{HD12}. The main difference is in the preparation of the initial models for the \cmfgen\ computations. When modeling the radiative transfer for SNe, we need massive star models evolved from the main sequence until iron-core collapse together with a computation of the explosion using radiation hydrodynamics. Here, our initial conditions are taken from the  simulations of high-velocity collisions of red-giant stars performed by \citet{ryu_collisions_23}.  Two stars in the vicinity of a supermassive BH and with a high relative velocity come into collision, leading to the formation of a double-shock structure crossing each stellar envelope while the denser He-core of each component continues on its trajectory largely  unimpeded. The huge dissipation of energy during the collision produces debris that expand quasi-spherically at supersonic speeds, very much like what is produced in SN explosions. Once the dynamical phase is over, which takes less than a day for red-giant star collisions, all mass shells have essentially the same radius-to-velocity ratio $R/V=t$, where $t$ is the time since the onset of the collision; all mass shells move ballistically and the expansion is homologous (any interaction that may take place with the interstellar medium is ignored in this work). The remapping into \cmfgen\ is then straightforward and practically equivalent to the approach used for SNe II (see, for example, \citealt{HD19}).

Our sample is composed of eight models from the red-giant star collisions presented in \citet{ryu_collisions_23}. We take the \arepo\ simulations at about 1\,d and build angular averages at all radii for the density and the temperature. In all models, the total ejecta mass is 60--80\% of the sum of the mass of each colliding star because their denser He core remains unaffected by the collision -- only the lower density stellar envelopes are shocked and modeled in this work while the two He-core travelling within the ejecta in opposite directions have a negligible impact on the observables. The composition of the material that makes up the ejecta is essentially primordial and uniform. For simplicity, we adopt a solar composition in this work. We force the velocity to exactly match $R/t$, where $t$ is the average value of $R/V$ in the original, 1-D model.  In Fig.~\ref{fig_init}, we show the initial spherical-averaged ejecta density and temperature profiles for our model set (we truncated the outer part interacting with the ambient medium in the \arepo\ simulations). The density structure is similar in the four models with the highest collision velocity (i.e., 10000\,\kms) and the smallest impact parameter (i.e., 0.04\,$R_\star$). In those four models ejecta masses cover the range 1.19 to 1.59\,\msun\ but the kinetic energy per unit mass is the same within a few percent. In contrast, the temperature in those four models is greater in the more extended progenitors (i.e., rs100 versus rs10). This temperature offset arises from the greater cooling from expansion in more compact progenitors (see, e.g., Sect.~4.1 of \citealt{DH10}), which is the primary effect at the origin of the light curve contrast between blue-supergiant (BSG) and red-supergiant (RSG) star explosions (i.e., SNe II-pec like SN\,1987A and SNe II-Plateau). The other four models with a greater impact parameter (i.e., 0.4 or 0.8\,$R_\star$) or smaller collision velocity (i.e., $v_{\rm rel}$ of 2500 or 5000\,\kms) yield weaker energetics and have more complicated density and temperature profiles. For example, the maximum density and temperature in model rs10/v1e4/b0p8 occurs well above the inner ejecta layers. In the weakest collision rs10/v2p5e3/b0p04, the kinetic energy is only 2.6$\times$10$^{49}$\,erg, which is a factor of ten smaller than in model rs100/v1e4/b0p04 and its radiative energy content is 100 times smaller. This different radiative energy content at 1\,d is the root cause of the differences in radiative properties throughout the photospheric phase (i.e., at times when the ejecta are optically thick). Our model nomenclature is such that model rs100/v1e4/b0p04 stands for the collision model of two stars with a surface radius of 100\,\rsun\ colliding at a relative velocity of 10000\,\kms\ and with an impact parameter of 0.04\,$R_\star$. A summary of the properties of our model set is provided in Table~\ref{tab_init}.

  At 1\,d, these models are remapped into the NLTE time-dependent radiative transfer code \cmfgen\ \citep{HD12} and evolved until they become optically thin at 20--40\,d. A solar-metallicity composition is assumed for all species. Specifically, we treat H, He, C, N, O, Ne, Na, Mg, Al, Si, S, Ar, Ca, Sc, Ti, Cr, Fe, Co, and Ni. We use the updated atomic data described in \citet{blondin_21aefx_23} and include the following ions for the above species: H\one, He\one-\two, C\one--\four, N\one--\five, O\one--\six, Ne\two--\five, Na\one, Mg\two--\five, Al\two--\three, Si\two--\five, S\two--\five, Ar\one--\three, Ca\two--\five, Sc\two--\three, Ti\two--\three, Cr\two--\six, Fe\one--\six, Co\two--\six, and Ni\one--\six. With our model atom, we treat 24000 levels and a total of 1.8 million bound-bound transitions. These are mostly coming from iron-group elements, in particular Fe, Co, and Ni (and of these three elements, Fe is the most important followed by Ni). We assume all isotopes are stable and thus ignore any radioactive decay and associated non-thermal processes. The grid employs 80 grid points uniformly spaced on a logarithmic optical-depth scale. Remapping the grid is necessary at the start of every timestep as well as during the calculation in order to track recombination fronts, if present. To cover the full evolution, a total of about 35 timesteps were computed in all models except for the weaker collisions. The main \cmfgen\ output of interest for this study is the emergent flux, which is calculated at each time step in the observer's frame from the far-UV to the far-IR. This spectrum can be shown directly or used to compute the bolometric luminosity $L_{\rm bol}$ or the absolute magnitude in various filters. We selected the Swift $UVW2$ and $UVW1$ filters and the optical $U$ and $V$ filters. The $UVW1$ filter is a close analogue of the NUV filter that will be on board ULTRASAT \citep{ULTRASAT}.  

For the weaker collisions, the models had to be stopped at 5--10\,d because of convergence difficulties associated with a steep recombination front, high densities, and low temperatures. Hence, for the three models rs10/v1e4/b0p8, rs10/v5e3/b0p04, and rs10/v2p5e3/b0p04, the full evolution computed by \cmfgen\ is truncated (i.e., stopped before the ejecta are optically thin). In order to provide complete bolometric light curves for the full set of models, these three models were computed with the grey radiation-hydrodynamics code \voned\ \citep{livne_93,dlw10a,dlw10b}, and in that case the simulation proceeded successfully until 50\,d.

    Since the present calculations share similarities with core-collapse SNe, we include comparisons to the type II-Plateau SN model X of \citet[named RSG here]{bostroem_22acko_23} and the type II-peculiar SN model a4 of \citet[named BSG here]{DH_2pec_19}. Model RSG (BSG) provides a close match to SN\,2022acko (SN\,1987A; photometric data from \citet{hamuy_87A_88} is shown for this SN in Fig.~\ref{fig_photometry}). We choose this Type II-Plateau model because it is one of the few of that type that has been calculated with \cmfgen\ at times prior to 10\,d, hence it is better suited for comparison to these fast evolving stellar collisions. The RSG model corresponds to an ejecta of 8.16\,\msun\ with a kinetic energy  of 0.6\,$\times$\,10$^{51}$\,erg. The BSG model corresponds to an ejecta of 13.22\,\msun\ and kinetic energy  1.26\,$\times$\,10$^{51}$\,erg. The same color coding is used for all models throughout the paper.
    
 \begin{figure}
\centering
\includegraphics[width=\hsize]{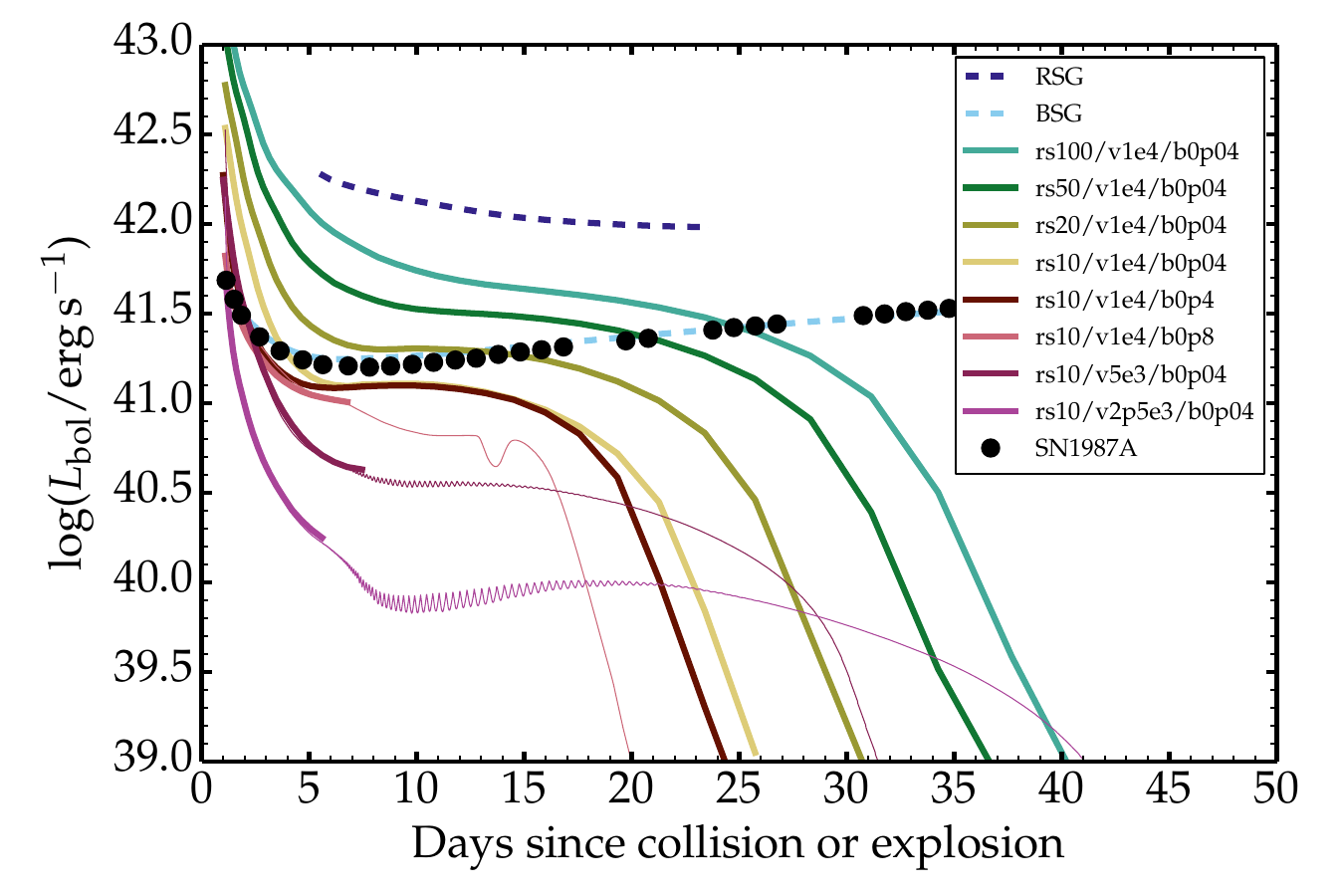}
\caption{Bolometric light curves of red-giant star collisions computed with \cmfgen. We include results for our set of eight models as well as those for the BSG explosion model a4 from \citet{DH_2pec_19} and the RSG explosion model X from \citet{bostroem_22acko_23}. We add the inferred values for SN\,1987A \citep{hamuy_87A_88}. For the three faintest models, the \cmfgen\ light curve (thick line) is completed with the \voned\ results (thin line of the same color).
\label{fig_lbol}
}
\end{figure}

\begin{figure}[tb!]
\centering
\includegraphics[width=\hsize]{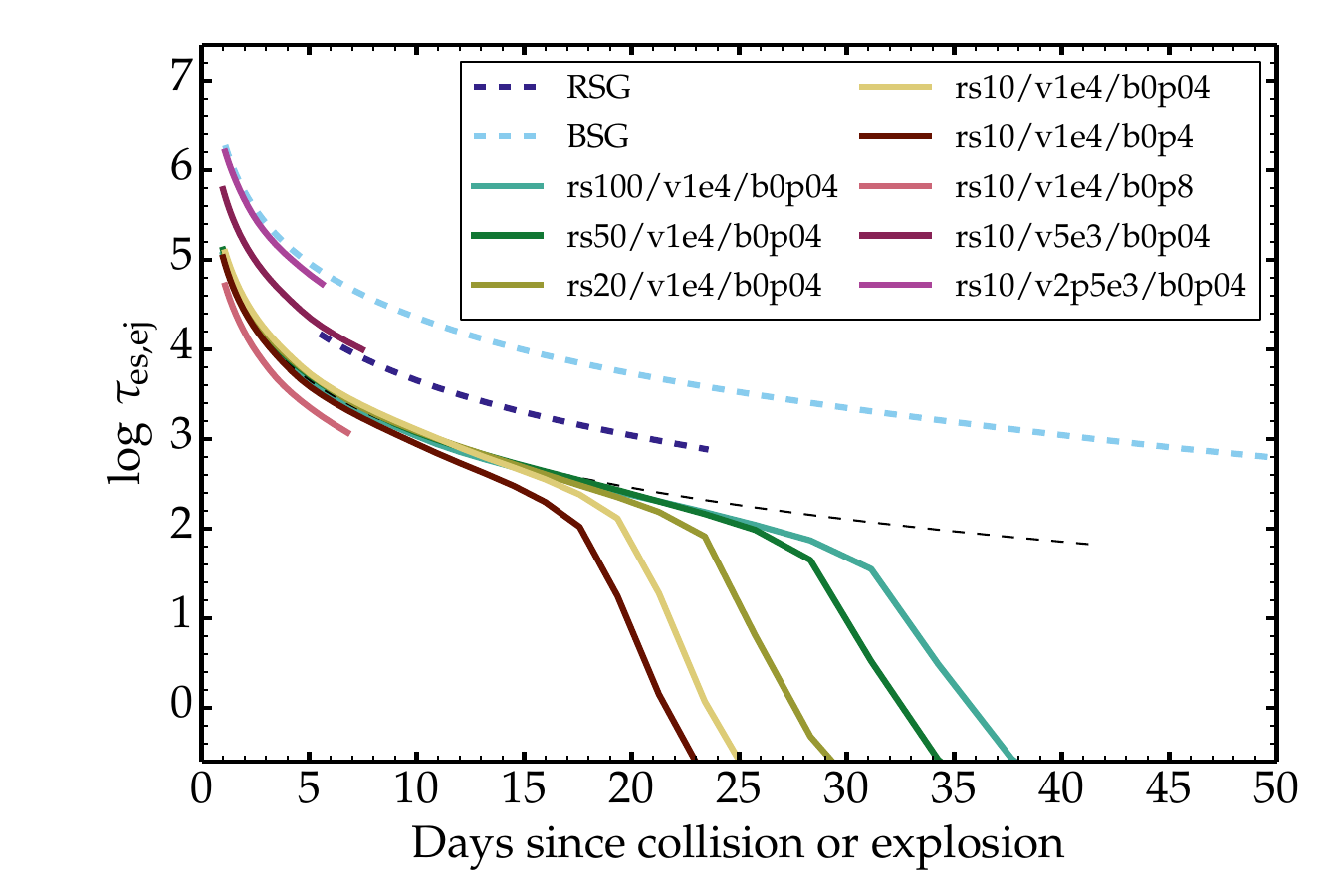}
   \caption{Same as Fig.~\ref{fig_lbol} but now showing the evolution of the total electron-scattering optical depth. The thin black dashed line indicates a $1/t^2$ slope expected at constant ionization.
}
\label{fig_tau_es}
\end{figure}

 \section{Results}
 \label{sect_res}

Figure~\ref{fig_lbol} shows the bolometric light curve computed with \cmfgen\ for our set of eight red-giant star collisions (as discussed above, the bolometric light curves for the three faintest models is completed after 5--10\,d by the results obtained with the grey radiation hydrodynamics code \voned). All simulations show a steeply declining luminosity at 1\,d (approximately following $t^{-1.4}$ to $t^{-2.2}$) for a maximum value that was probably well in excess of 10$^{43}$\,\ergs\ (the peak is expected to occur at the time of collision; \citealt{ryu_collisions_23}).  The decline rate is greater for more compact giants (rs10 versus rs100), larger impact parameters (b0p8 versus b0p04) and weaker collision velocities (v2p5e3 versus v1e4). After about 5\,d, the bolometric light curves flatten and enter a plateau phase that is less luminous as we progress to weaker collisions or more compact progenitors (the range covers from about 10$^{41.7}$ down to 10$^{40}$\,\ergs). The plateau in brightness occurs during the recombination phase, when the photospheric temperature is around 6000\,K, and arises from a rough balance between the radial expansion of the ejecta and the recession in mass space of the photosphere.  The duration of the plateau phase is set by a combination of factors, including the amount of stored radiative energy (see Table~\ref{tab_init}), the rate at which this stored energy is released through the receding photosphere, and the optical depth of the ejecta (which depends on mass, expansion rate, temperature, composition or progenitor radius) -- all these quantities are not independent. The duration of this plateau phase, which ends at the same time as the optically-thick phase (Fig.~\ref{fig_tau_es}), is shorter for more compact progenitors for the same collision parameters (impact parameter and velocity; models rs100/, rs50/, rs20/, and rs10/v1e4/b0p04). The main reason is the reduced stored radiative energy at 1\,d in collisions of more compact giants. In the weaker collisions, the expansion rate of the debris is smaller and contributes to lengthening the optical-thick phase, although the model is underluminous throughout that photospheric phase. For example, models rs100/v1e4/b0p04 and rs10/v2p5e3/b0p04 have roughly the same optically-thick phase duration but over that time, the former radiates 50 times more energy (it started at 1\,d with 140 times more stored radiative energy). Unlike core-collapse SNe, there are no unstable isotopes to provide a persistent power source so once the debris from stellar collisions turn optically thin, the power drops precipitously. In practice, the ejecta would reprocess the radiation from the two He-cores of the colliding stars, which are embedded within the ejecta, and the luminosity would level off at a rate of $10^{35}-10^{37}$\,\ergs\ (progenitors rs10 or rs100). When comparing with the BSG and RSG explosion models, the main ingredient causing the change in bolometric luminosity is the change in progenitor radius because of the impact it has on expansion cooling (see, for example, the correlations established in the context of SNe by \citealt{popov_93}). Another ingredient is radioactive decay, which is at the origin of the steady slow brightening in the BSG explosion model.

\begin{figure*}
   \centering
    \begin{subfigure}[b]{0.45\textwidth}
       \centering
       \includegraphics[width=\textwidth]{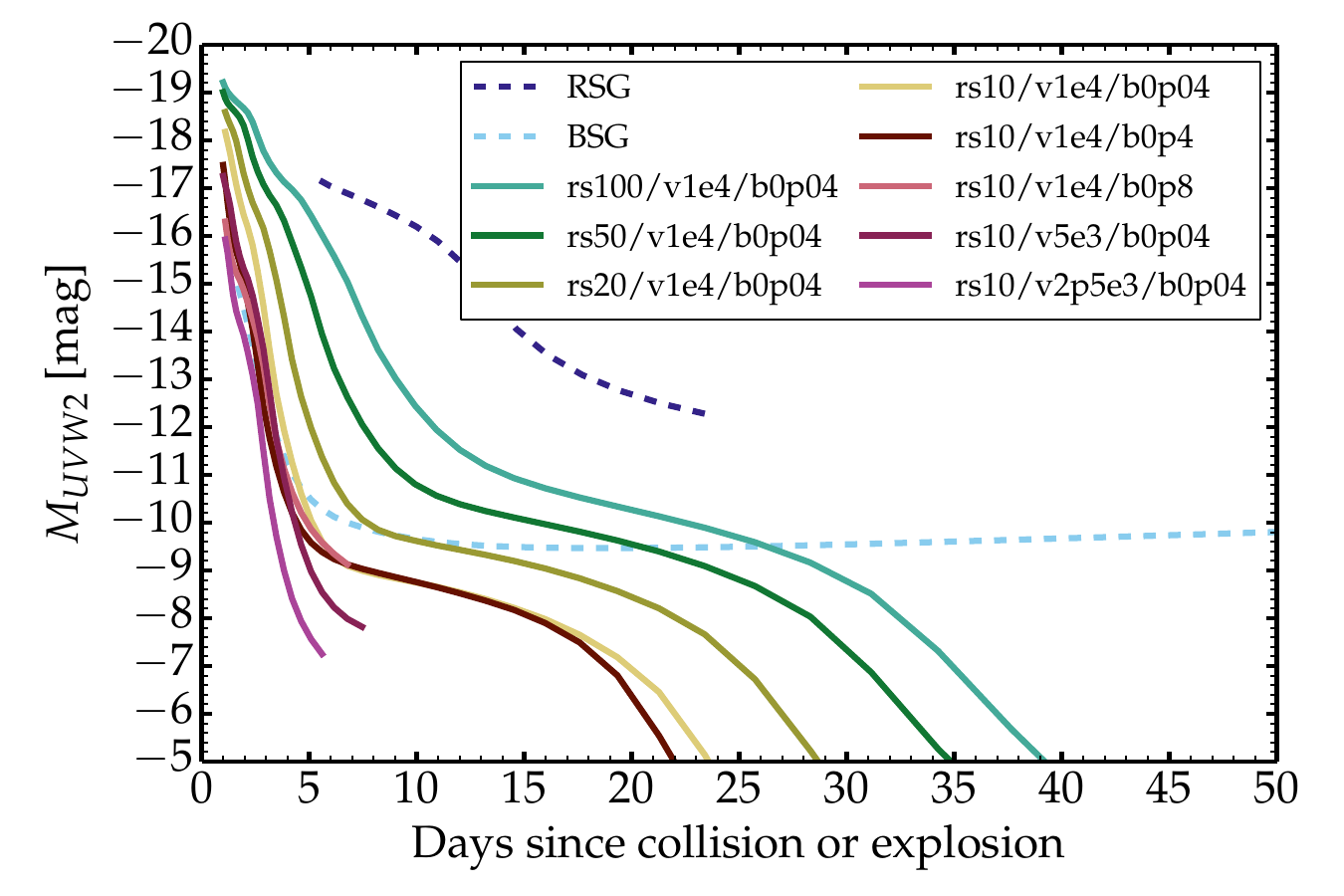}
    \end{subfigure}
    \hfill
    \centering
    \begin{subfigure}[b]{0.45\textwidth}
       \centering
       \includegraphics[width=\textwidth]{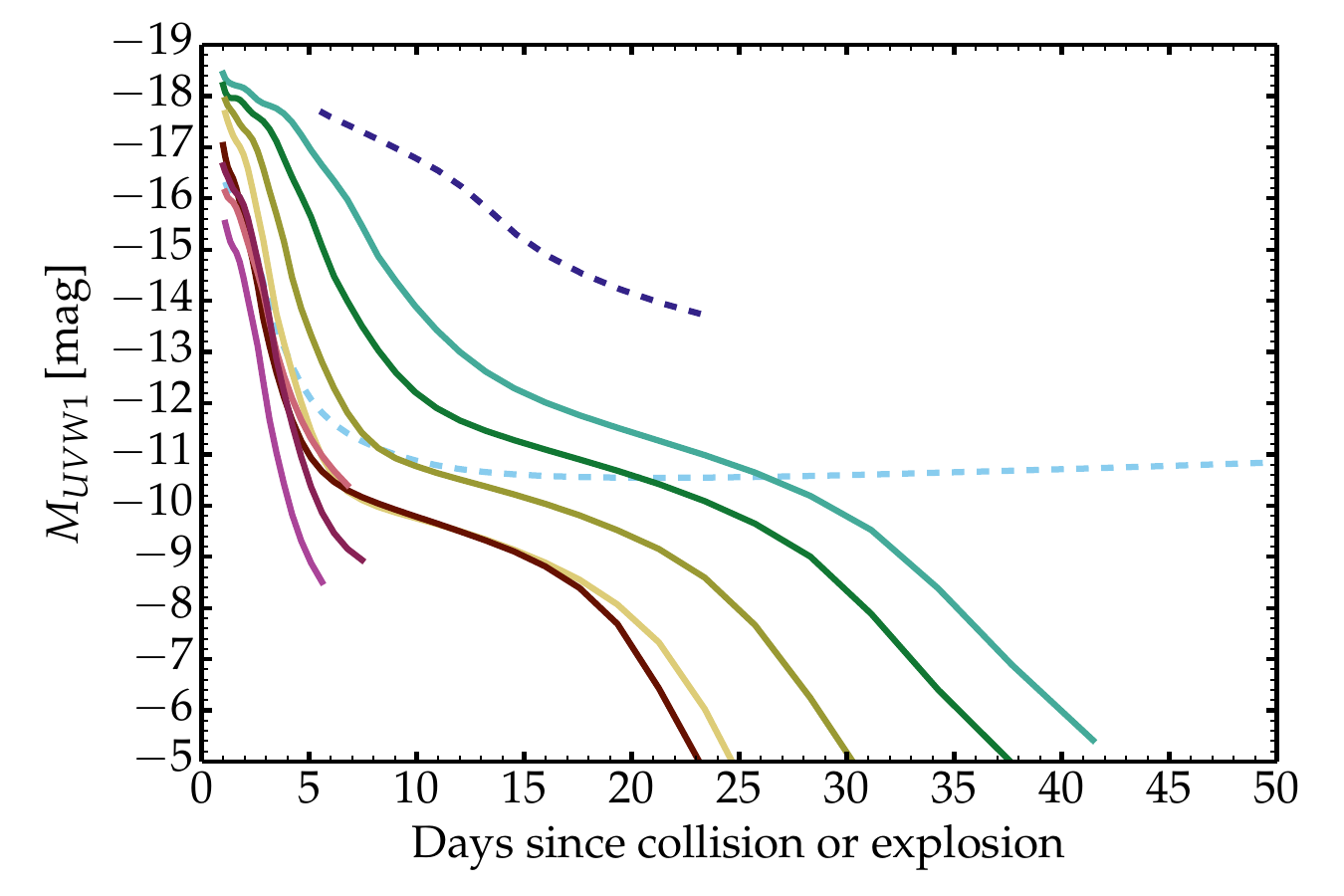}
    \end{subfigure}
    \vskip\baselineskip
   \centering
    \begin{subfigure}[b]{0.45\textwidth}
       \centering
       \includegraphics[width=\textwidth]{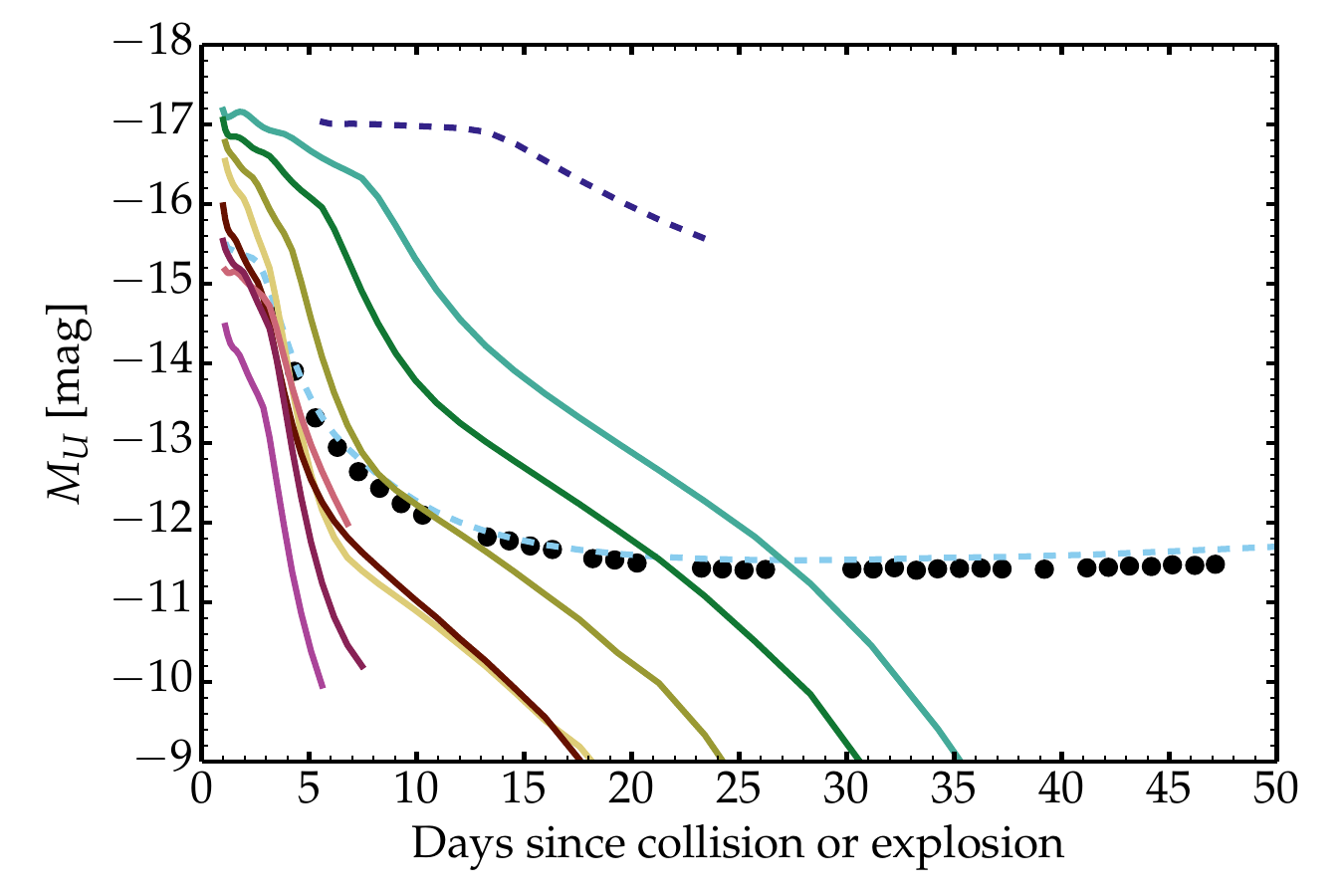}
    \end{subfigure}
    \hfill
    \centering
    \begin{subfigure}[b]{0.45\textwidth}
       \centering
       \includegraphics[width=\textwidth]{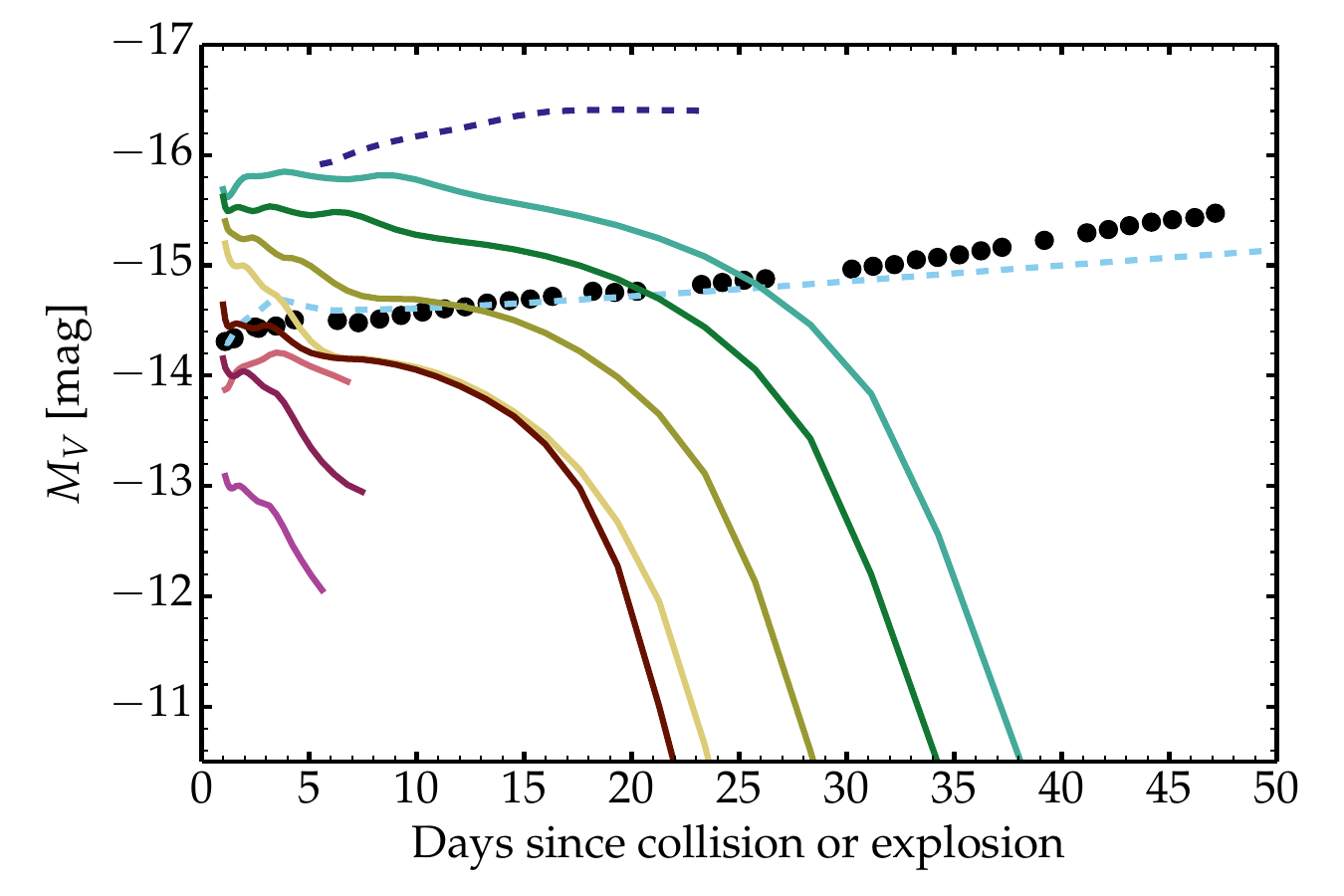}
    \end{subfigure}
\caption{Same as Fig.~\ref{fig_lbol}, but now showing the photometric properties for the filters $UVW2$, $UVW1$, $U$, and $V$ (data for SN\,1987A are shown for the optical filters). The light curves for the three models corresponding to weaker collisions are truncated because of convergence difficulties with \cmfgen\ during the recombination phase.
\label{fig_photometry}
}
\end{figure*}

Figure~\ref{fig_photometry} shows the photometric properties computed with \cmfgen\ for our model set. 
In the UV (filters $UVW2$ and $UVW1$), all light curves show a similar behavior as obtained for the bolometric luminosity, with a decline from a maximum of $-18$ to $-19$\,mag down to $-8$ to $-10$\,mag at 5--15\,d. The UV plateau lasts until the debris turn optically thin. At that time the UV brightness plunges again. In contrast, in the BSG explosion model, radioactive decay maintains the UV brightness on a faint plateau. In the RSG explosion model, the much larger progenitor radius (i.e., $\sim$\,500\,\rsun) causes the UV luminosity to remain large for much longer as a result of the reduced cooling from expansion. It is clear, however, that all models are UV luminous for at least a few days after the collision.

In the $U$ band, all light curves show a continuous drop without any obvious plateau, while in the $V$ band, the light curves are essentially a continuous plateau throughout the optically-thick phase until they plunge when the debris become optically thin. This behavior of the $U$ band, also shared by UV filters, is caused by the low temperature (which systematically drops, although more slowly after the onset of the recombination phase) and the enhanced metal-line blanketing in the blue part of the optical and in the UV -- see below). The $U$ and $V$-band light curves for the high-velocity collisions (models rs10/, rs20/, and rs50/v1e4/b0p04) are at the same brightness level as obtained for the BSG explosion model and observed for SN\,1987A but differ in shape due to the lack of radioactive decay. The difference is in part associated with the lower ejecta mass (on the order of 1--2\,\msun) compared to $\gtrsim$\,10\,\msun\ for the BSG model and SN\,1987A \citep{DH_2pec_19}.

\begin{figure}
\centering
\includegraphics[width=0.8\hsize]{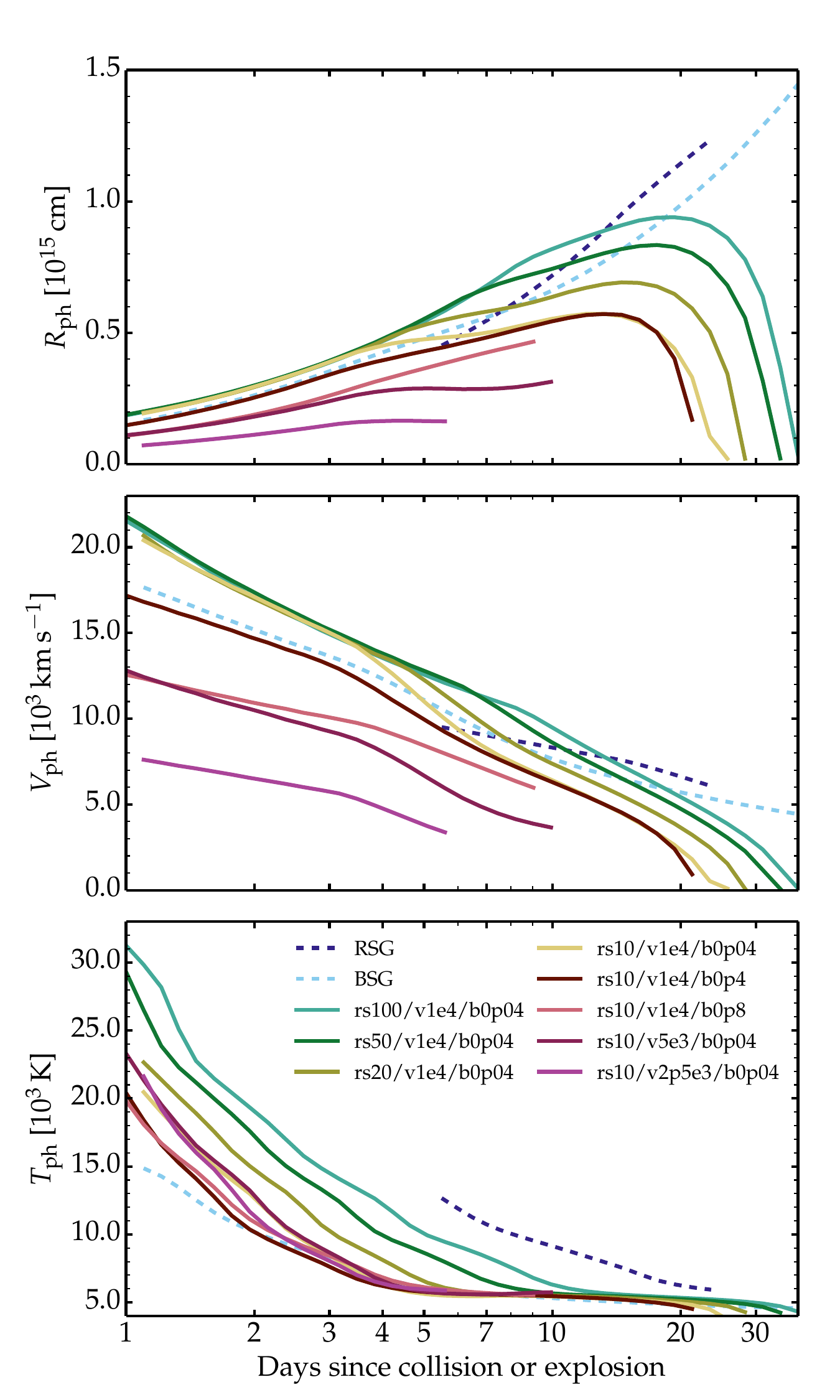}
   \caption{Evolution of the photospheric radius (top), velocity (middle) and gas temperature (bottom) for our model set. For comparison, we also show the results for the BSG and RSG explosion models. The photosphere is taken at the location where $\tau_{\rm es}=2/3$.
}
\label{fig_phot}
\end{figure}

Figure~\ref{fig_phot} shows the photospheric properties of our model set. Photospheric radii cover from about 0.1 to a maximum of 0.5 to 0.9$\times$\,10$^{15}$\,cm, the greater for the bigger progenitors or more energetic collisions. The photospheric velocity follows the same evolution for the four violent collisions (i.e., $v_{\rm vel}=$\,10000\,\kms) with maximum values of 20000\,\kms\ at 1\,d followed by a rapid and steady decline. A steeper drop occurs after about a week; it is more pronounced in the more compact progenitors because of the earlier recombination, causing the photosphere to recede faster in mass, and therefore in velocity space. In the weaker collisions, the reduced kinetic energy in the ejecta leads to maximum velocities at 1\,d of about 7000 to 17000\,\kms. Although the \cmfgen\ calculations are truncated, the evolution of the photospheric velocity is expected to follow that of other models. The photospheric temperatures cover the range 18--23\,kK at 1\,d, dropping steadily until reaching the H-recombination temperature between 4 to 12\,d. The onset of the recombination phase occurs earlier in weaker explosion and more compact progenitors. During that recombination phase, all models have exactly the same photospheric temperature of 5000--6000\,K.

\begin{figure*}
   \centering
    \begin{subfigure}[b]{0.45\textwidth}
       \centering
       \includegraphics[width=\textwidth]{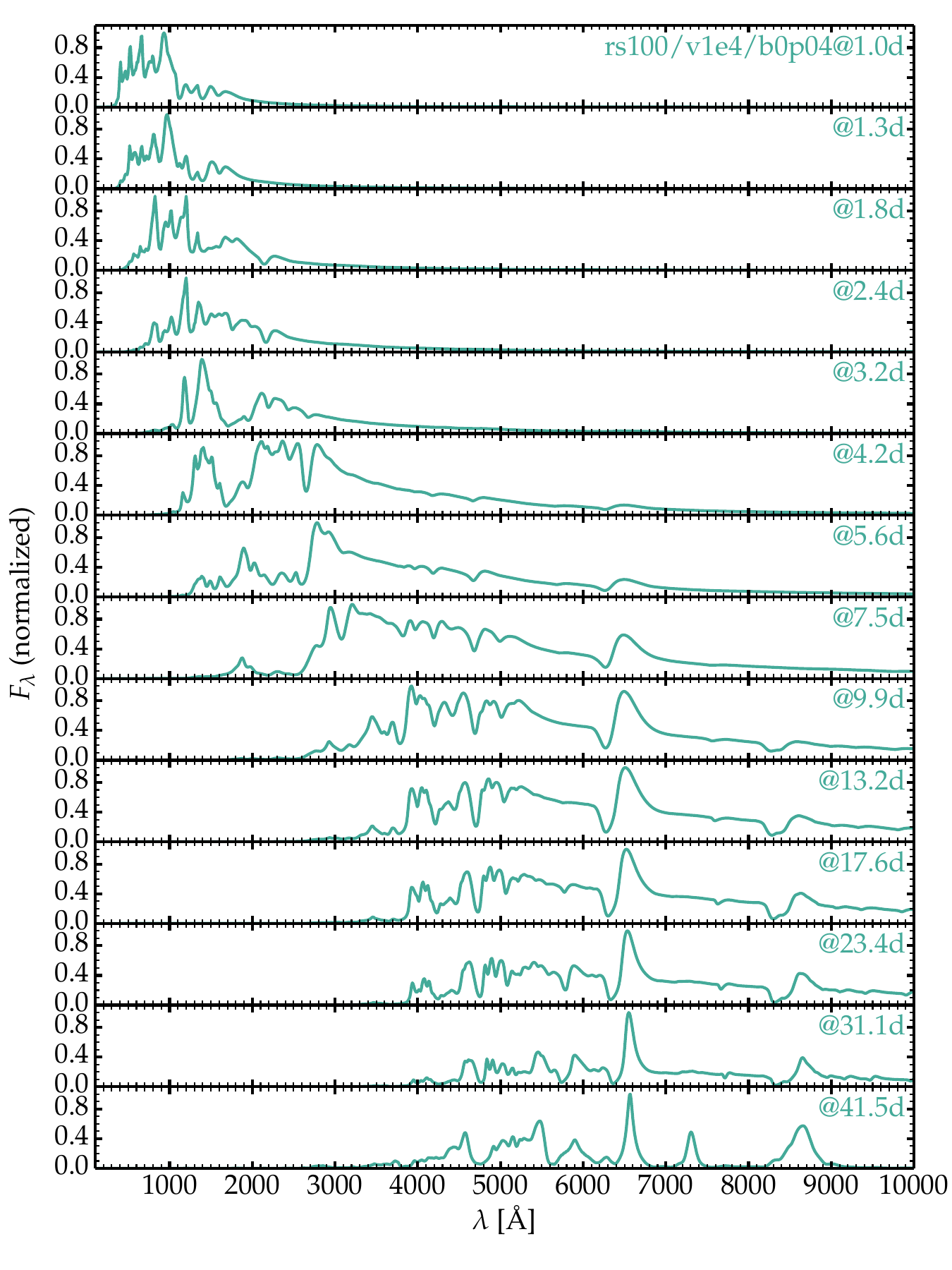}
    \end{subfigure}
    \hfill
    \centering
    \begin{subfigure}[b]{0.45\textwidth}
       \centering
       \includegraphics[width=\textwidth]{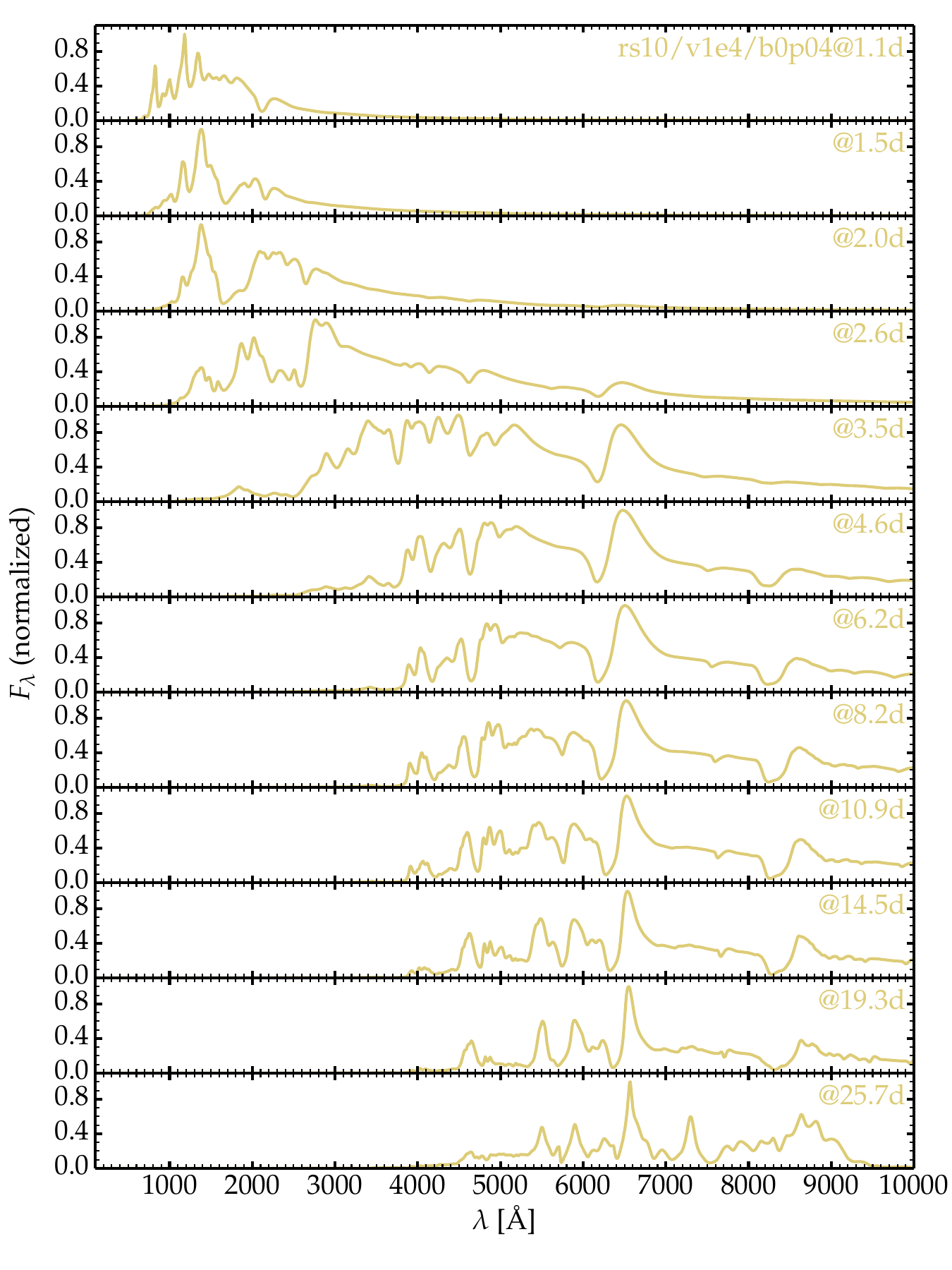}
    \end{subfigure}
\caption{Spectral evolution for models rs100/v1e4/b0p04 (left) and rs10/v1e4/b0p04 (right). Although the time step adopted in our \cmfgen\ simulations is 10\% of the current time, we show spectra here with an increment of about 30\% of the current time (i.e., epochs shown are 1, 1.3, 1.8\,d etc). All spectra are normalized to a maximum flux of unity.}
\label{fig_spec_seq}
\end{figure*}

\begin{figure}
\centering
\includegraphics[width=0.9\hsize]{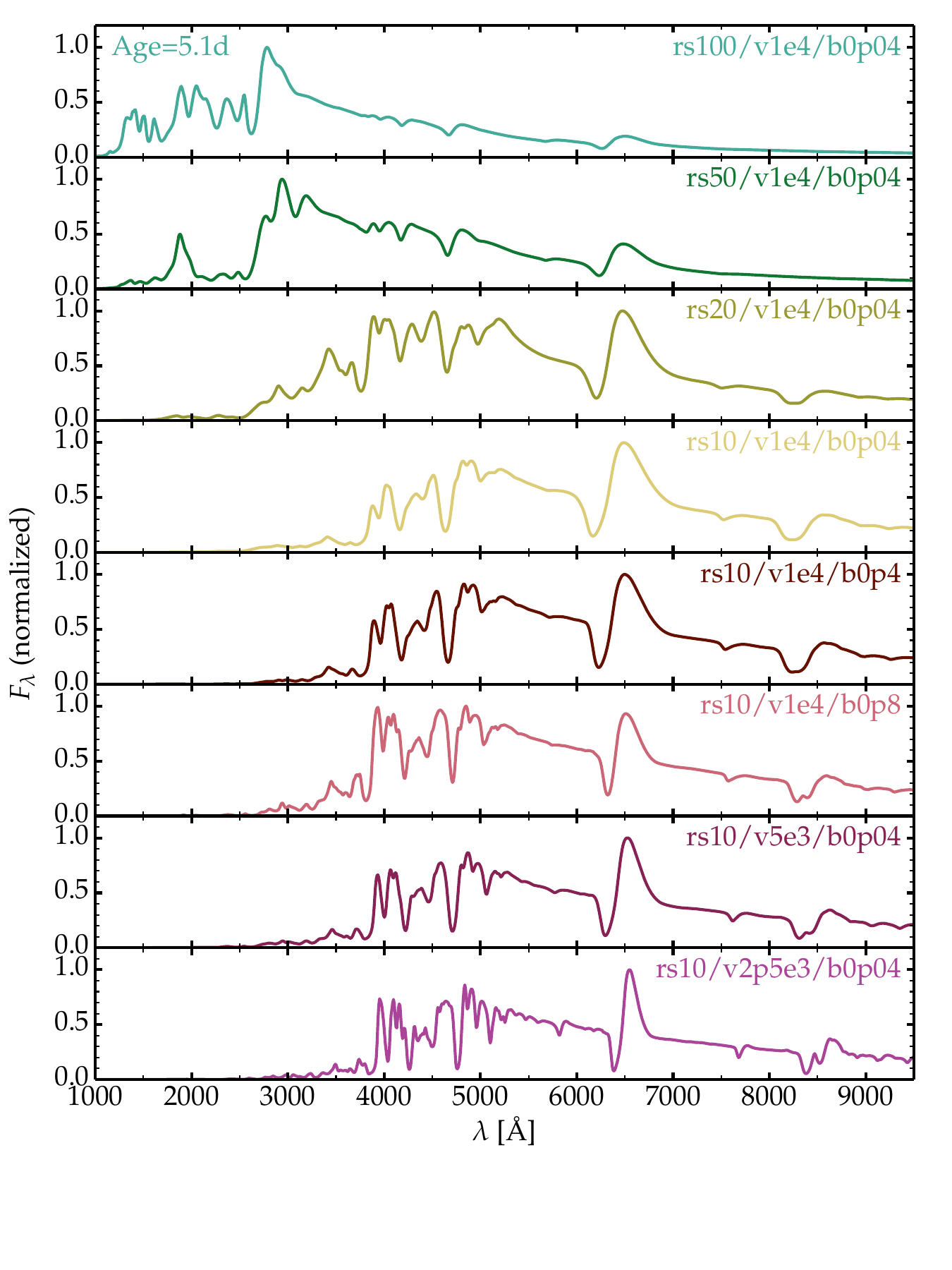}
\vspace{-1.cm}
\caption{Spectral comparison covering the UV and the optical  for our model set at 5.1\,d after collision. All spectra are normalized to a maximum flux of unity (the difference in luminosity between models is shown in Fig.~\ref{fig_lbol}).
\label{fig_comp_at_5d}
}
\end{figure}

Figure~\ref{fig_spec_seq} illustrates the spectral evolution for stellar-collisions models rs100/v1e4/b0p04 and rs10/v1e4/b0p04 (a presentation of spectra for the entire model set is given in the appendix, in Figs.~\ref{fig_spec_seq_appendix1}--\ref{fig_spec_seq_appendix2}). The spectra, which are shown from 100\,\AA\ to 1\,$\mu$m, are normalized to the maximum flux within that range. The luminosity and photometric properties can be gathered from Figs.~\ref{fig_lbol} and \ref{fig_photometry}. The spectral evolution reflects the color evolution from UV bright to optically bright, with a transition that occurs later in the rs100 model. Because of the large expansion rate of these ejecta (and by extension of their photospheres and spectral-formation regions), the spectra exhibit only a few strong lines, either isolated ones from the H\one\ Balmer series (in particular H$\alpha$ and H$\beta$), or resonance transitions like \nad, or  blends of numerous transitions associated with metal species like Ti\two\ or Fe\two. Strong lines are also predicted in the rapidly changing UV range such as \mgiidoub\ and a forest of Fe\three\ lines, though mostly prior the recombination phase (see, for example, discussion in \citealt{bostroem_22acko_23}).  Metals such as  Ti\two\ or Fe\two\ contribute significant blanketing in the optical band during the recombination phase (i.e., once $T_{\rm ph}$ drops to 5000--6000\,K; Fig.~\ref{fig_phot}). These are responsible for the broad peaks and valleys between 4000 and 7000\,\AA. Apparent in the evolution is also the fact that the main qualitative distinction in the two sequences is the rate at which the SED shifts to the red. For the same color, the two models are nearly identical (e.g., model rs10/v1e4/b0p04 at 9.9\,d appears nearly identical to model rs100/v1e4/b0p04 at the earlier time of 4.6\,d). By extension, the similarity would extend also to the BSG explosion model or the observations of SN\,1987A. At early times, they would appear analogous to the present predictions for model rs50/v1e4/b0p04. 

Figure~\ref{fig_comp_at_5d} gives a more direct comparison between the spectra of our model set at a given time of 5.1\,d. The top four models in the figure mostly differ in that their photospheric temperature has not dropped as much (rs100 is still hot and ionized while rs10 is entering the recombination phase). The bottom four panel show much less contrast because they correspond to models in the recombination phase. The most obvious difference is the narrower lines of model rs10/v2p5e3/b0p04 which arise from the smaller expansion rate of the ejecta, itself caused by the much weaker collision relative to other models in our set. In this ejecta from a weak collision, both absorption and emission components of all lines are reduced in extent (most visibly in strong lines like H$\alpha$), and the associated reduction of line overlap makes individual line features appear sharper.

\section{Conclusion}
\label{sect_conc}

We have presented NLTE time-dependent radiative transfer calculations with \cmfgen\ for the debris resulting from high-velocity collisions of 1\,\msun\ red-giant stars expected to occur in dense, nuclear clusters. These complement the earlier estimates of \citet{ryu_collisions_23}, \citet{AmaroSeoane2023} and \citet{amaro_col_23b}. The \cmfgen\ calculations are based on the hydrodynamical simulations performed with \arepo\ by \citet{ryu_collisions_23}.  The remapping from \arepo\ to \cmfgen\ is taken at 1\,d after collision when the quasi-spherical ejecta are essentially in homologous expansion. This shocked stellar gas is reminiscent of supergiant star explosions following gravitational collapse and thus the results from our \cmfgen\ calculations are qualitatively similar to those obtained for Type II SNe, either from BSG \citep{DH_2pec_19} or RSG star progenitors \citep{HD19}. Our sample includes eight models varying in progenitor radius (10 to 100\,\rsun), collision velocity (2500, 5000, and 10000\,\kms), and impact parameter (0.04, 0.4, 0.8\,$R_\star$). In all cases, the two colliding stars have identical properties. These collisions yield ejecta with masses $M_{\rm ej}$ in the range 1.19--1.60\,\msun, kinetic energies $E_{\rm kin}$ in the range 0.26 up to 4.0$\times$\,10$^{50}$\,erg, and stored radiative energies at 1\,d in the range 0.01 up to 1.6$\times$\,10$^{49}$\,erg. 

In our model set, the ratio $E_{\rm kin}/M_{\rm ej}$ is in the range 0.16 to 2.5$\times$\,10$^{50}$\,erg/\msun, which brackets the value of our BSG explosion model, which has an $E_{\rm kin}/M_{\rm ej}$  of $\sim$\,10$^{50}$\,erg/\msun. Together with the comparable progenitor radius (10 to 100\,\rsun\ compared to 50\,\rsun\ for the BSG model), this explains to a large extent the very similar bolometric light curve, multi-band light curves, as well as spectral evolution. The reduced ejecta mass implies a much smaller ejecta optical depth and therefore a shorter photospheric phase. This is aggravated by the absence of unstable isotopes and associated radioactive decay heating. Stellar collisions never rebrighten significantly (in contrast to BSG explosions like SN\,1987A) and eventually plunge precipitously into oblivion. More specifically, our stellar collisions have bolometric luminosities on the order of 10$^{42}$--10$^{43}$\,\ergs\ at 1\,d, dropping to a plateau brightness of 10$^{40}$--10$^{41.7}$\,\ergs\ after 5--10\,d, and fading suddenly at the end of the optically-thick phase at 15--40\,d, depending on the progenitor radius and the strength of the collision. All simulations are UV bright (magnitude on the order of $-18$ to $-19$\,mag) for a few days and should be detectable by UV transient surveys like ULTRASAT  \citep{ULTRASAT} or UVEX \citep{uvex}.

The spectral evolution is also reminiscent of that observed for SN\,1987A or obtained for the BSG model  \citep{DH_2pec_19}. Our energetic stellar collisions span a similar range in photospheric velocity, radius, and temperature but scan that range much faster owing to the much smaller ejecta mass.  Weaker collisions yield results that are analogous to underenergetic  BSG explosions \citet[see, for example, model a3]{DH_2pec_19}, and are characterized by lower luminosities and narrower lines at all times.

In this work, we have neglected the potential power injection from interaction with ISM and fallback accretion into the supermassive BH. Interaction with ISM would yield a sustained bolometric luminosity until very late times, perhaps for years, depending on the ISM density and extent, although this power would most likely emerge in the UV \citep{DH_interaction_22,d23_interaction}. In addition, it would produce peculiar profiles. During the photospheric phase, broad boxy emission would be present on top of the strongest lines like H$\alpha$ and would thus modify the spectral properties at all times. Fallback accretion could also produce a sustained or delayed rebrightening depending on the history of this accretion (see, for example, \citealt{dexter_kasen_13}). It may produce super-luminous events, such as the peculiar type II SNe OGLE14-073 \citep{terreran_slsn2_17}  or iPTF14hls \citep{arcavi_iptf14hls,d18_iptf14hls}.

In this work, we have assumed a solar metallicity. Stars in dense nuclear clusters may form there and exhibit a solar or supra-solar metallicity. If they were formed outside the nuclear center, as may occur from globular cluster infall, they would likely have a sub-solar metallicity (see, e.g., \citealt{do_gc_15}, \citealt{schultheis_apogee_20}). Variations in metallicity would not alter the bolometric curve, but it would modulate the colors (the higher the metallicity, the greater the impact of metal-line blanketing at $<$\,5500\,\AA) and the spectra (they would exhibit varying strength in metal lines, both from the forests of lines due to  iron-group elements as well as from isolated transitions associated with Na\one\ or Ca\two). A discussion of these effects is presented in \citet{zpap}.

Considering their peak luminosity, which rivals that of tidal disruption events \citep[see][for a review]{gezari_tde_21} or Type II SNe, BDCs can appear as nuclear transients. It is possible that BDCs have already been detected by ongoing surveys like ASAS-SN or ZTF and will be detected by upcoming surveys such as ULTRASAT or LSST. The light curves and spectra presented in this paper will be instrumental in distinguishing BDCs from other types of nuclear transients observed in these surveys. While we focused on the observables of BDCs involving giant stars, in principle, BDCs involving stars of different types (e.g., main-sequence stars) can generate a luminous electromagnetic display as long as a significant portion of collision kinetic energy is converted into radiation. Investigations into the detection rate of such events and their observable characteristics among stars of various types will be pursued in follow-up projects.

\begin{acknowledgements}
This research was supported by the Munich Institute for Astro-, Particle and BioPhysics (MIAPbP) which is funded by the Deutsche Forschungsgemeinschaft (DFG, German Research Foundation) under Germany's Excellence Strategy  -- EXC-2094 -- 390783311. PAS acknowledges the funds from the ``European Union
NextGenerationEU/PRTR'', Programa de Planes Complementarios I+D+I (ref. ASFAE/2022/014). This work was granted access to the HPC resources of TGCC under the allocation 2022 -- A0130410554 made by GENCI, France. This research has made use of NASA's Astrophysics Data System Bibliographic Services.

\end{acknowledgements}


\appendix

\section{Additional figures for all models}

\begin{figure*}
    \centering
    \begin{subfigure}[b]{0.45\textwidth}
       \centering
       \includegraphics[width=\textwidth]{montage_rs100_v1e4_b0p04_0p9d_list_UV_optical.pdf}
    \end{subfigure}
    \hfill
    \centering
    \begin{subfigure}[b]{0.45\textwidth}
       \centering
       \includegraphics[width=\textwidth]{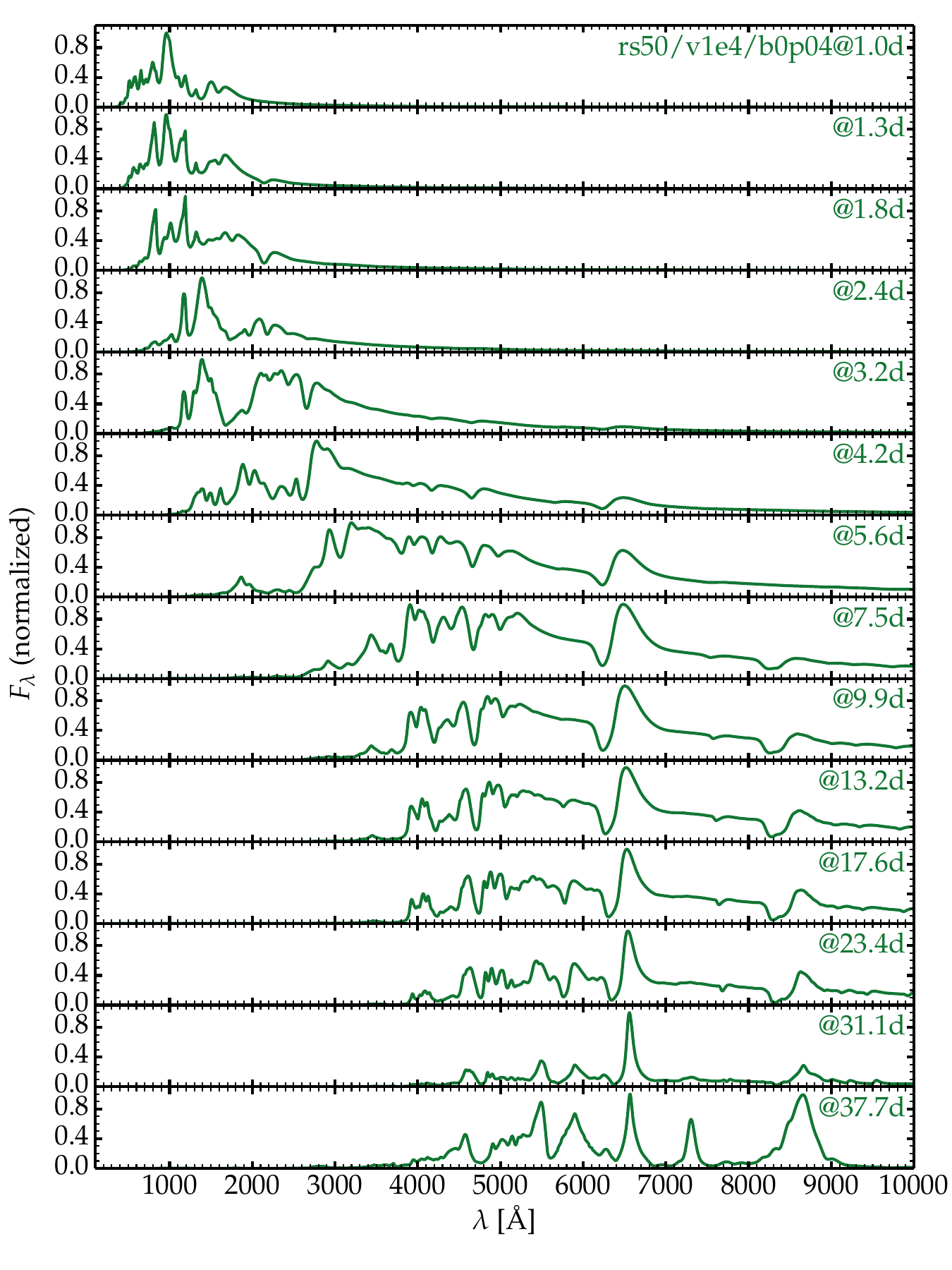}
    \end{subfigure}
    \vskip\baselineskip
    \centering
    \begin{subfigure}[b]{0.45\textwidth}
       \centering
       \includegraphics[width=\textwidth]{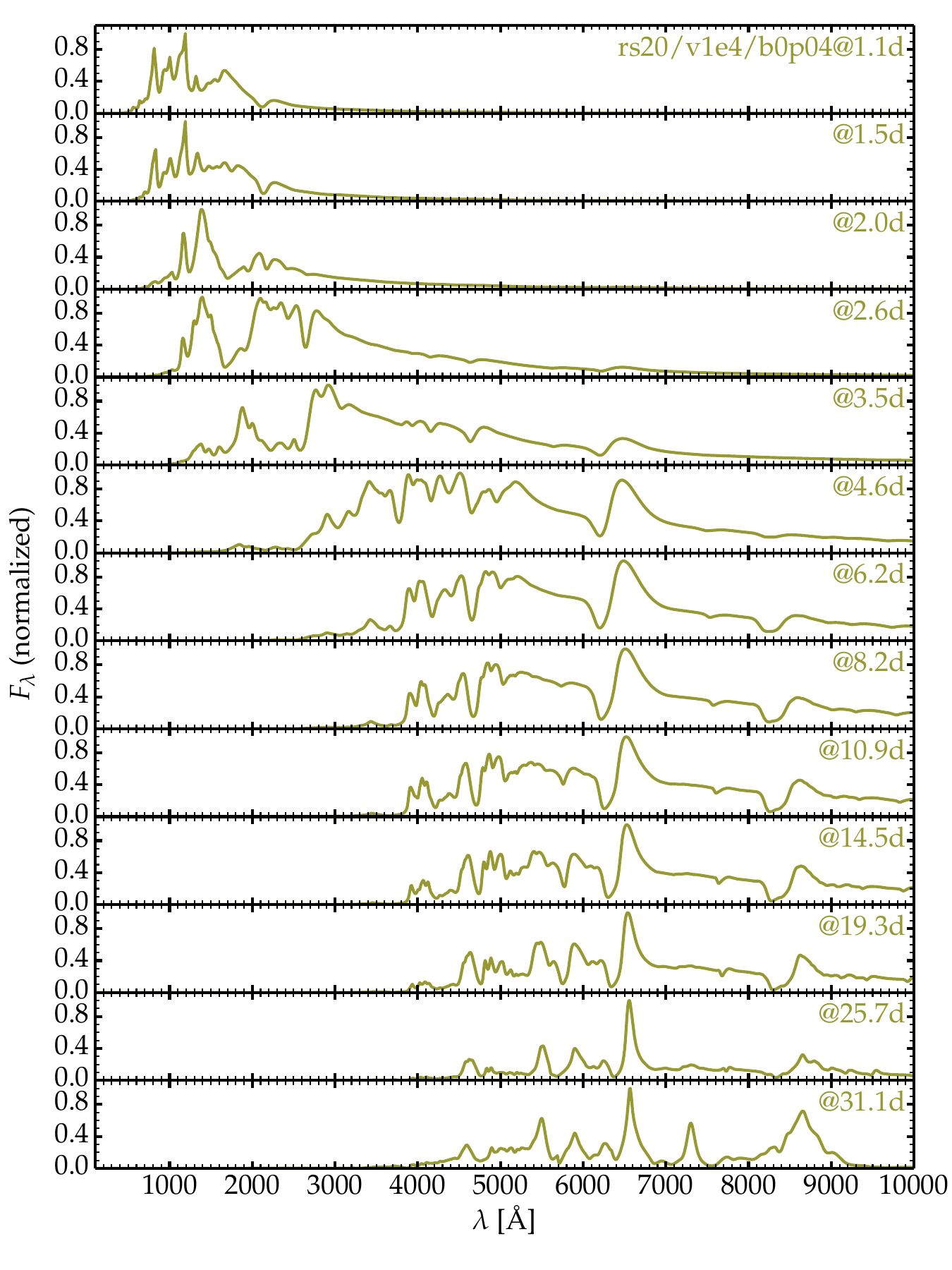}
    \end{subfigure}
    \hfill
    \centering
    \begin{subfigure}[b]{0.45\textwidth}
       \centering
       \includegraphics[width=\textwidth]{montage_rs10_v1e4_b0p04_1p0d_list_UV_optical.pdf}
    \end{subfigure}
    \caption{Spectral evolution for models rs100/v1e4/b0p04 (top left), rs50/v1e4/b0p04 (top right), rs20/v1e4/b0p04 (bottom left), and rs10/v1e4/b0p04 (bottom right). The main difference between these models is the red-giant star radius (100, 50, 20, and 10\,\rsun, given in the same order).}
\label{fig_spec_seq_appendix1}
\end{figure*}

\begin{figure*}
    \centering
    \begin{subfigure}[b]{0.45\textwidth}
       \centering
       \includegraphics[width=\textwidth]{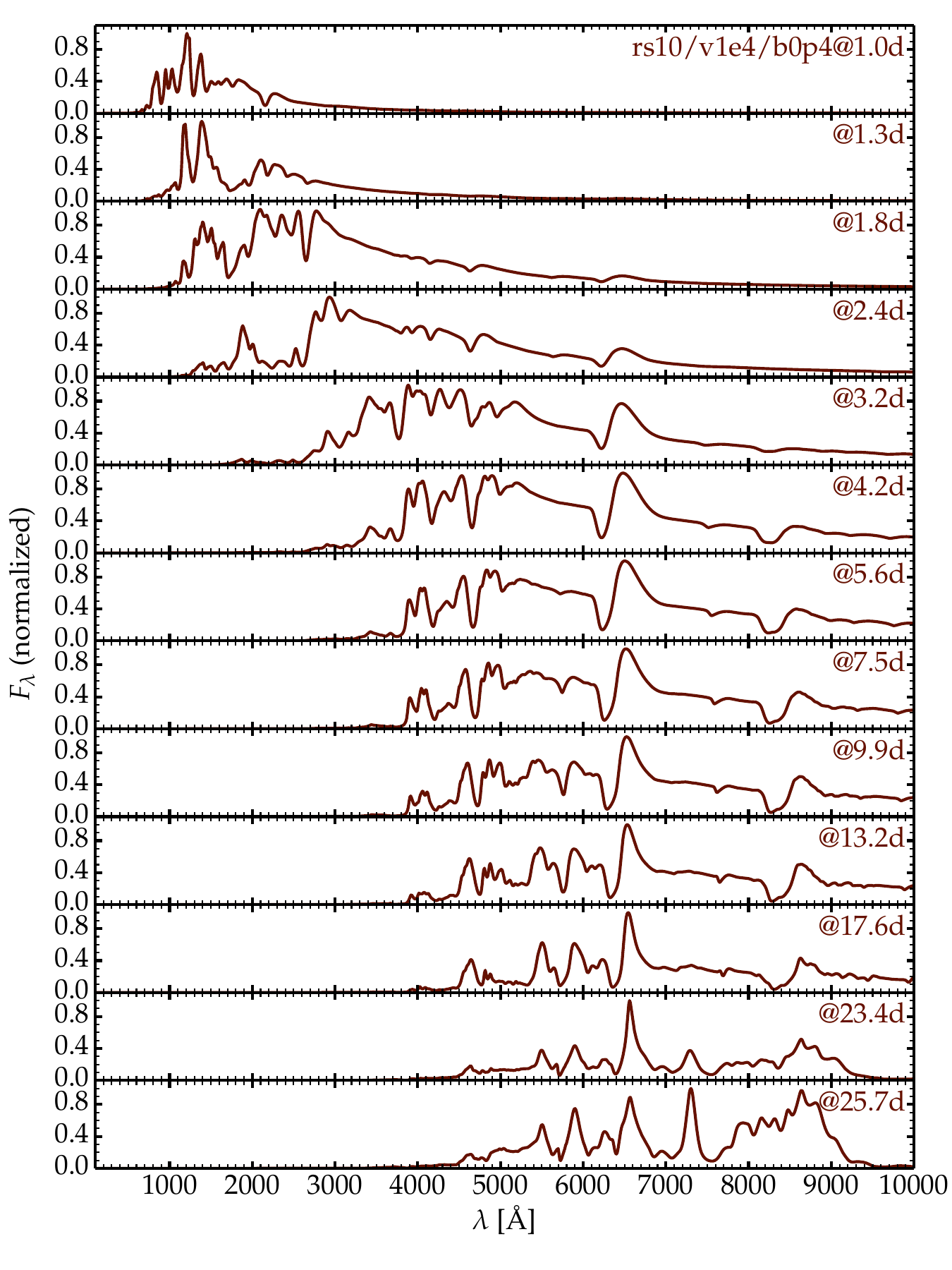}
    \end{subfigure}
    \hfill
    \centering
    \begin{subfigure}[b]{0.45\textwidth}
       \centering
       \includegraphics[width=\textwidth]{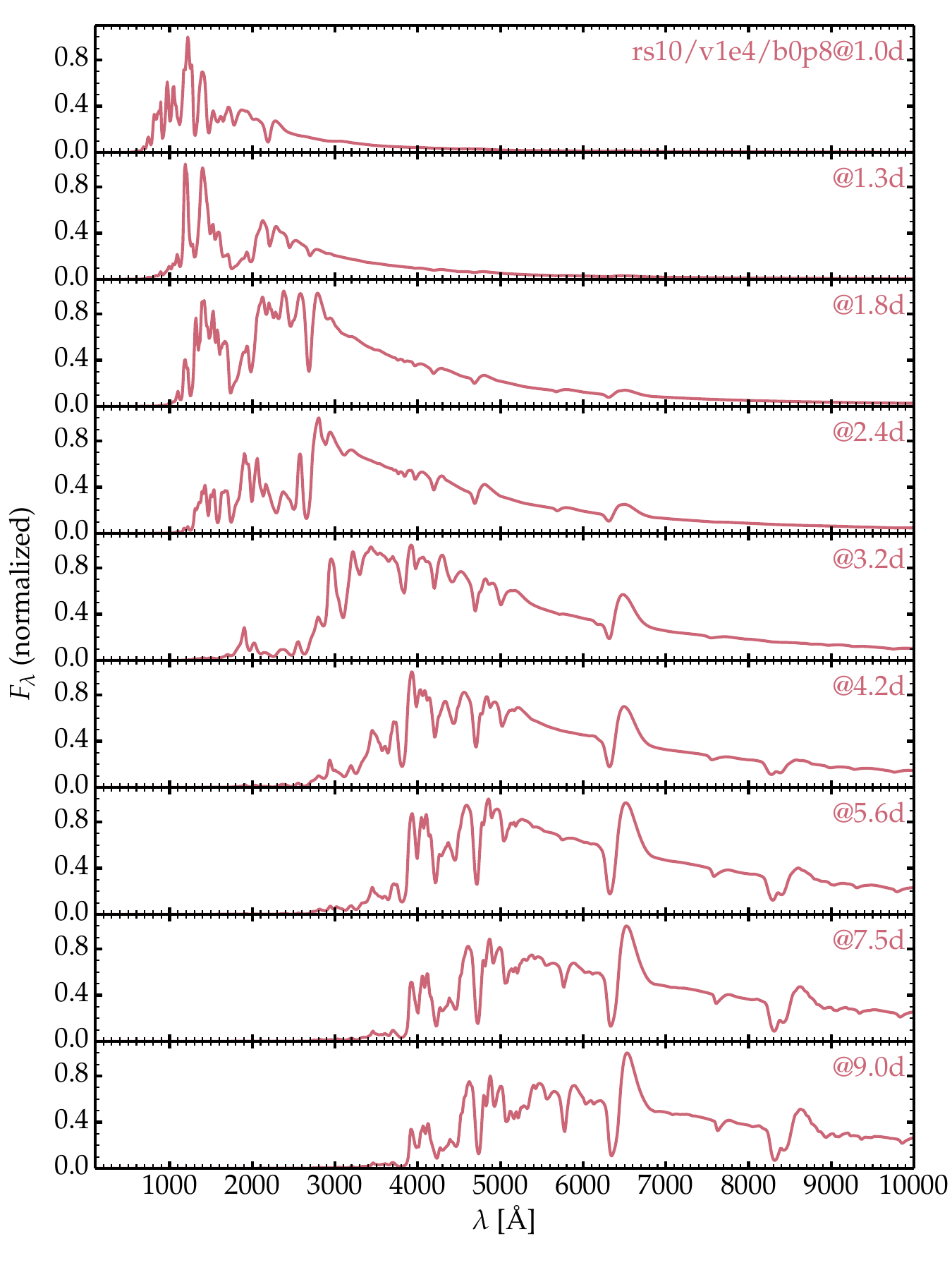}
    \end{subfigure}
    \vskip\baselineskip
    \centering
    \begin{subfigure}[b]{0.45\textwidth}
       \centering
       \includegraphics[width=\textwidth]{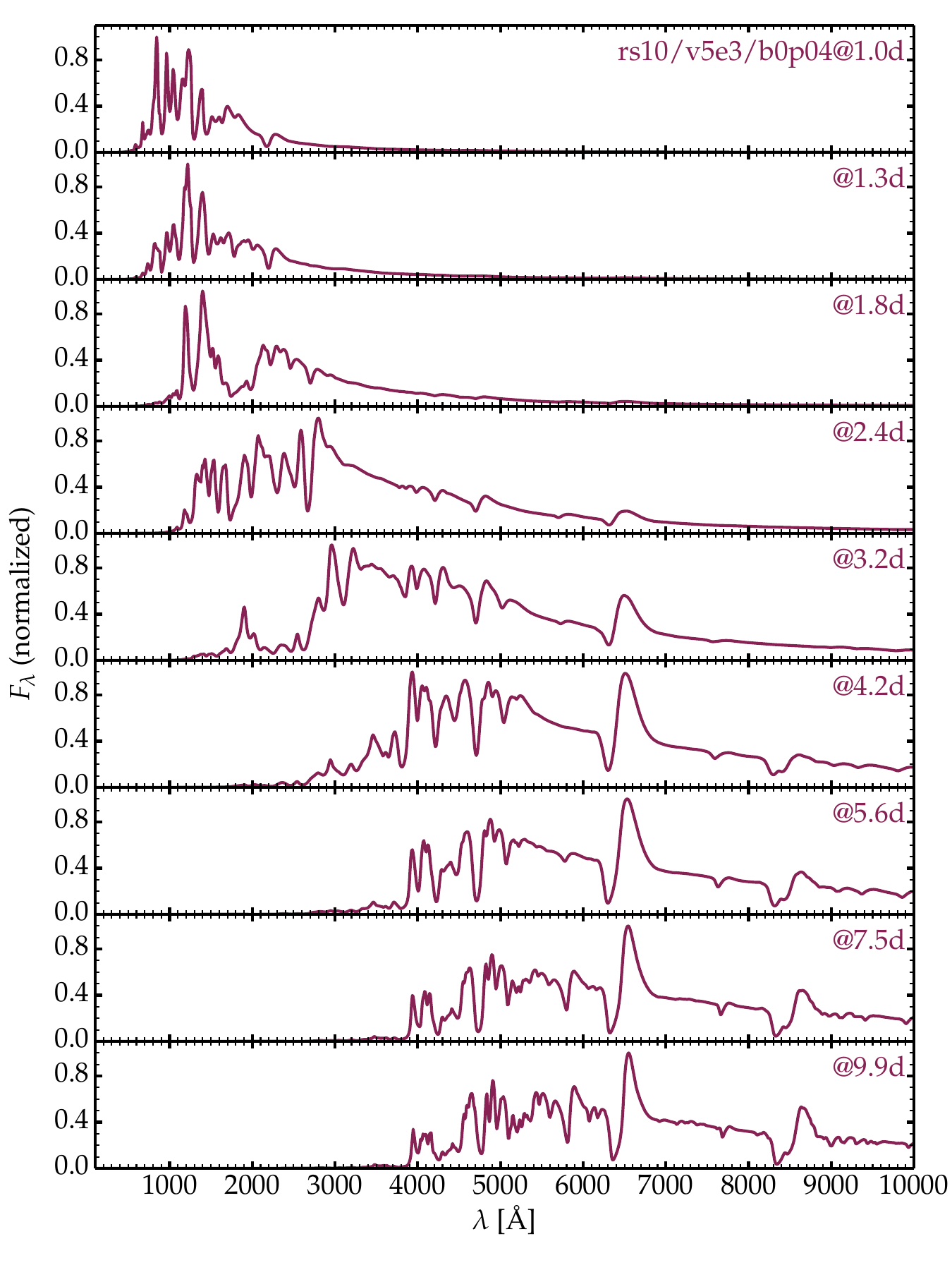}
    \end{subfigure}
    \hfill
    \centering
    \begin{subfigure}[b]{0.45\textwidth}
       \centering
       \includegraphics[width=\textwidth]{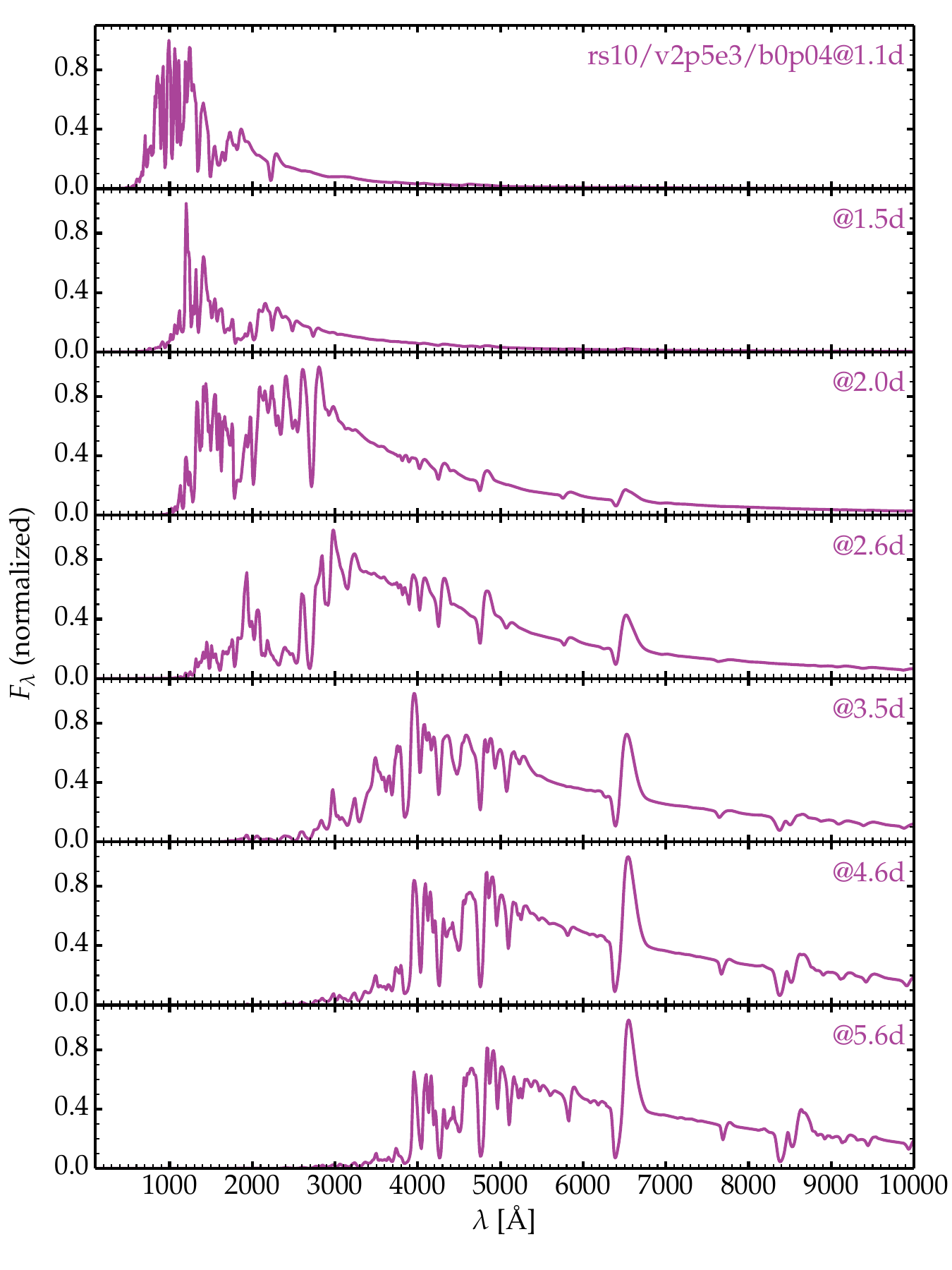}
    \end{subfigure}
    \caption{Spectral evolution for models rs10/v1e4/b0p4 (top left), rs10/v1e4/b0p8 (top right), rs10/v5e3/b0p04 (bottom left), and rs10/v2p5e3/b0p04 (bottom right). The main difference between these models is the impact parameter (0.4 or 0.8\,$R_\star$; top row) or the velocity of the collision (5000 or 2500\,\kms; bottom row). For the last three models, the \cmfgen\ calculation is truncated at 5--10\,d; convergence difficulties prevented the calculations until the ejecta were fully optically thin (see Fig.~\ref{fig_tau_es}).}
\label{fig_spec_seq_appendix2}
\end{figure*}

\end{document}